\documentclass[superscriptaddress, nofootinbib, longbibliography, amsmath, amssymb, twocolumn]{revtex4-2}

\usepackage[english]{babel}
\usepackage[T1]{fontenc}
\usepackage[utf8]{inputenc}
\usepackage{graphicx}
\usepackage{hyperref}
\usepackage{booktabs}
\usepackage{xspace}
\usepackage{lmodern}
\usepackage{microtype}
\usepackage{amsmath}
\usepackage{amssymb}
\usepackage{mathtools}
\usepackage{enumitem}
\usepackage{tabularx}
\usepackage{float}
\hypersetup{
    hidelinks,
    colorlinks=true,
    breaklinks=true,
    urlcolor=blue,
    allcolors=blue,
    pdftitle={Recent advances in the Self-Referencing Embedding Strings (SELFIES) library},
    pdfauthor={Alston Lo, Robert Pollice, AkshatKumar Nigam, Andrew D. White, Mario Krenn and Alán Aspuru-Guzik}
}

\newcommand*{\SMILES}{{S}{\footnotesize MILES}\xspace}
\newcommand*{\DeepSMILES}{{D}{\footnotesize EEP}{S}{\footnotesize MILES}\xspace}
\newcommand*{\SELFIES}{{S}{\footnotesize ELFIES}\xspace}

\newcommand*{\code}[1]{\textcolor{tqblue}{\texttt{#1}}}
\newcommand*{\selfieslib}{\code{selfies}\xspace}

\newcommand*{\bX}{\mathbf{X}}

\newcommand*{\bS}{\mathbf{S}}

\usepackage{listings, xcolor}
\definecolor{codegreen}{rgb}{0,0.6,0}
\definecolor{codegray}{rgb}{0.5,0.5,0.5}
\definecolor{codepurple}{rgb}{0.58,0,0.82}
\definecolor{tqblue}{HTML}{08293d}
\definecolor{backcolour}{HTML}{fefdf5}

\lstdefinestyle{pythonstyle}{
    backgroundcolor=\color{backcolour},   
    commentstyle=\color{codegreen},
    keywordstyle=\color{magenta},
    numberstyle=\tiny\color{codegray},
    stringstyle=\color{codepurple},
    basicstyle=%
        \ttfamily
        \color{tqblue}
        \lst@ifdisplaystyle\footnotesize\fi,
    breakatwhitespace=false,         
    breaklines=true,
    postbreak=\mbox{\textcolor{magenta}{$\hookrightarrow$}\space},
    captionpos=b,                    
    keepspaces=true,                 
    numbers=left,                    
    numbersep=5pt,                  
    showspaces=false,                
    showstringspaces=false,
    showtabs=false,                  
    tabsize=2
}

\DeclareMathOperator{\atomtype}{\textsf{type}}
\newcommand{\valence}{\nu}
\newcommand{\concat}{\,}

\newcommand{\atomsymb}{\code{[} \concat \beta \concat \alpha \concat \code{]}}
\newcommand{\branchsymb}{\code{[} \concat \beta \concat \code{Branch} \concat \ell \concat \code{]}}
\newcommand{\ringsymb}{\code{[} \beta \concat \code{Ring} \concat \ell \concat \code{]}}
\newcommand{\stereoringsymb}{\code{[} \concat \beta_1 \concat \beta_2 \concat \code{Ring} \concat \ell \concat \code{]}}

\newcommand{\latestversion}{\code{2.1.1}\xspace}

\begin{document}

\title{Recent advances in the Self-Referencing Embedding Strings (SELFIES) library}

\author{Alston Lo}
\email{alston.lo@mail.utoronto.ca}
\affiliation{Department of Computer Science, University of Toronto, Canada.}

\author{Robert Pollice}
\email{robert.pollice@gmail.com}
\affiliation{Department of Computer Science, University of Toronto, Canada.}
\affiliation{Chemical Physics Theory Group, Department of Chemistry, University of Toronto, Canada.}

\author{AkshatKumar Nigam}
\affiliation{Department of Computer Science, Stanford University, California, USA.}

\author{Andrew~D.~White}
\affiliation{Department of Chemical Engineering, University of Rochester, USA.}

\author{Mario Krenn}
\affiliation{Max Planck Institute for the Science of Light (MPL), Erlangen, Germany.}

\author{Al\'an Aspuru-Guzik}
\email{alan@aspuru.com}
\affiliation{Department of Computer Science, University of Toronto, Canada.}
\affiliation{Chemical Physics Theory Group, Department of Chemistry, University of Toronto, Canada.}
\affiliation{Vector Institute for Artificial Intelligence, Toronto, Canada.}
\affiliation{Canadian Institute for Advanced Research (CIFAR) Lebovic Fellow, Toronto, Canada.}

\begin{abstract}
    String-based molecular representations play a crucial role in cheminformatics applications, and with the growing success of deep learning in chemistry, have been readily adopted into machine learning pipelines. However, traditional string-based representations such as \SMILES are often prone to syntactic and semantic errors when produced by generative models. To address these problems, a novel representation, SELF-referencIng Embedded Strings (\SELFIES), was proposed that is inherently 100\% robust, alongside an accompanying open-source implementation \selfieslib. Since then, we have generalized \SELFIES to support a wider range of molecules and semantic constraints and streamlined its underlying grammar. We have implemented this updated representation in subsequent versions of \selfieslib, where we have also made major advances with respect to design, efficiency, and supported features.
    Hence, we present the current status of \selfieslib (version \code{2.1.1}) in this manuscript. Our library, \selfieslib{}, is available at GitHub (\href{https://github.com/aspuru-guzik-group/selfies}{https://github.com/aspuru-guzik-group/selfies}).
\end{abstract}

\maketitle
\newpage
\tableofcontents

\section{Introduction}
\label{sec:introduction} 

In recent years, machine learning (ML) has become a powerful tool to tackle challenging problems in chemistry. Machine learning pipelines involve three crucial elements: data, representation, and models. Choosing the proper representation is important as it defines the space of available models available to work with the data, as well as impacting directly model performance. For molecules, one of the more widely-used classes of representations encode molecules as strings (i.e., the string-based molecular representations). These representations are popular since they can leverage the rich collection of ML tools that have been developed for sequential data \cite{warr2011representation, wigh2022review}. Historically, the most employed string representation is the Simplified Molecular Input Line Entry System (\SMILES), which was introduced by Weininger in 1988~\cite{weininger1988smiles}. Currently, \SMILES has become the \textit{de facto} standard representation in cheminformatics and has historically been a key component of central applications in the field, such as chemical databases. The main appeal of \SMILES is its simple underlying grammar, which allows for the rigorous specification of molecules in a manner that can be parsed efficiently, and which is readable for humans at least for small molecules.

However, in an ML setting, this grammar can carry two intrinsic weaknesses. First, many strings constructed from \SMILES symbols are \textit{syntactically} invalid due to the \SMILES grammar, i.e., the strings cannot be interpreted as molecular graphs \cite{gomez2018automatic,sanchez2018inverse}. For instance, \SMILES requires open and closing brackets to appear in matching pairs, so the \SMILES string \texttt{C(CC} is invalid. This is problematic because ML models that produce \SMILES strings, especially generative models, can be prone to these syntactic errors, rendering a significant fraction of their output meaningless. One strategy is to constrain the ML architecture to reduce the number of invalid structures, which has been demonstrated successfully in the literature \cite{kusner2017grammar, olivecrona2017molecular, popova2018deep}. This approach, of course, needs significant computational effort and cannot be transferred directly to other systems without model retraining, model architecture adjustments, or domain-specific design considerations. An alternative and more fundamental solution is to define representations that are inherently robust. A first step towards this direction was taken by \DeepSMILES~\cite{o2018deepsmiles}, a string-based representation derived from \SMILES that reworked some of its most syntactically susceptible rules. While \DeepSMILES solves most of the syntactical errors, it does not address the second weakness of \SMILES, namely, that even syntactically valid strings may not necessarily correspond to a physical molecule. Typically, this occurs when a string represents a molecular graph that exceeds normal chemical valences, in which case we call the string \textit{semantically} invalid. For example, the \SMILES string \texttt{CO=C} is semantically invalid because it erroneously specifies a trivalent oxygen atom, which is chemically unstable and reactive.

To eliminate both syntactic and semantic invalidities in string-based molecular representations on a fundamental level, an entirely new representation termed SELF-referencIng Embedded Strings (\SELFIES) has been proposed by some of us \cite{krenn2020self}. By construction, \SELFIES is 100\% \textit{robust} to both syntactic and semantic errors. That is, any combination of \SELFIES symbols specifies a molecular graph that obeys chemical valences. This is achieved through a small Chomsky type-2, context-free grammar \cite{hopcroft2006automata} that is augmented with self-referencing functions to handle the generation of branches and rings.
Since its release, \SELFIES has enabled or improved numerous applications, ranging from molecular design \cite{nigam2021beyond, shen2021deep, thiede2022curiosity, eckmann2022limo} to interpretability \cite{wellawatte2022model} to image-to-string and string-to-string translations \cite{rajan2020decimer, rajan2021stout}, and has been extended to incorporate functional groups and other fragments \cite{cheng2022group}. For an extensive summary of its applications and opportunities, we refer readers to the recent community paper on \SELFIES \cite{krenn2022selfies}.

Herein, we introduce \selfieslib \latestversion, the latest version of the open-source Python implementation of \SELFIES. In particular, we provide a detailed look into its history, developments, underlying algorithms, design, and performance. Together with the community, we have recently overviewed potential extensions and formulated 16 concrete future projects for \SELFIES and other robust molecular string representations \cite{krenn2022selfies}. We hope that this manuscript will also help in developing some of these extensions and ideas. Our software package \selfieslib can be installed with \code{pip} \code{install} \selfieslib and is available at GitHub
(\href{https://github.com/aspuru-guzik-group/selfies}{https://github.com/aspuru-guzik-group/selfies}) under the Apache 2.0 license, along with comprehensive documentation and tutorials.

\begin{table*}[ht]
    \setlist[itemize]{left=0pt, before={\begin{minipage}[t]{\hsize}}, after={\end{minipage}}}
	\caption{A timeline of the various releases of \selfieslib.}
	\vspace{\baselineskip}
	\centering
	\renewcommand*{\arraystretch}{1.3} 
	\begin{tabularx}{0.95\textwidth}{ccX}
	\toprule
	Version  & Year(s)             &  Description
	\\ 
	\midrule
	\code{0.1.1} & (Jun) 2019 &
	\begin{itemize} 
	\item Initial release of \selfieslib.
	\end{itemize}
	\hrule height 0pt
	\\
	\code{0.2.4} & (Oct) 2019 &
	\begin{itemize} 
	\item Release of \selfieslib that implements the representation from \citet{krenn2020self}.
	\end{itemize}
	\hrule height 0pt
	\\
	\code{1.0.x} & 2020-21 &
	\begin{itemize}  
	    \item Expanded the support of \selfieslib to a greater subset of \SMILES strings, including strings with aromatic atoms, isotopes, charged species, and certain stereochemical specifications. To do so, the underlying grammar used by \selfieslib was both streamlined and generalized. 
	    \item Added support for the customization of the semantic constraints used by \selfieslib.
	    \item Significantly improved the efficiency of translation between \SELFIES and \SMILES.
	    \item Added a variety of utility functions to make the handling of \SELFIES strings convenient.
	\end{itemize}  
	\\
	\code{2.0.x} &  2021 &
	\begin{itemize}
	    \item Updated the \SELFIES alphabet to be more human-readable and standardized. 
	    \item Improved handling of stereochemical specifications in \SELFIES involving ring bonds. 
	\end{itemize} 
	\\
    \code{2.1.x} & 2022 &
    \begin{itemize}
        \item Added support for explaining translations between \SELFIES and \SMILES through attributions. 
    \end{itemize} 
    \hrule height 0pt
    \\
	\bottomrule
	\end{tabularx}
	\label{tab:timeline}
\end{table*}

\section{Timeline and Advances}

The \selfieslib library version that implemented the representation from \citet{krenn2020self} was first released as \selfieslib \code{0.2.4} in 2019. This older version provided an API of two translation functions where a restricted subset of organic, uncharged, nonaromatic \SMILES strings could be converted to and from \SELFIES strings. In addition, the internal algorithms behind \selfieslib relied heavily on direct string manipulations, so they were computationally inefficient and difficult to maintain.
Since then, \selfieslib has undergone several major redesigns that have significantly advanced the algorithmic handling of both \SMILES and \SELFIES. Most importantly, the underlying grammar of \selfieslib has been streamlined and generalized in subsequent versions. We will now describe the changes up until \selfieslib \latestversion, the most recent version of \selfieslib at the time of publication of this work.

One major modification we made is that \selfieslib now uses directed molecular graphs to internally represent \SMILES and \SELFIES strings. This has afforded \selfieslib greater efficiency and flexibility and enabled a number of additional extensions to be made. For example, we added support for aromatic molecules by kekulizing \SMILES strings with aromatic symbols before they are translated into \SELFIES. Furthermore, we handle species with partial charges, radicals, and molecules with explicit hydrogens, non-standard isotopes, and stereochemical definitions in a fully syntactically and semantically robust way. Besides the standard constraints for the number of valences, users can now specify their own constraints and we provide built-in relaxed and stricter constraint presets that can be selected conveniently. Most recently, we introduced the ability to  trace the connection between input and output tokens when translating between \SELFIES and \SMILES. \autoref{tab:timeline} gives a brief changelog of the major releases of \selfieslib and their associated advancements.

While the ideas outlined in \citet{krenn2020self} ensure the validity of the representation remains at the core of \selfieslib, the manifold implementation improvements and extensions are the novelties that we detail in this paper. Hereafter, unless specified otherwise, we will use \selfieslib to refer to \selfieslib \latestversion in particular and \SELFIES to refer to the representation that \selfieslib \latestversion implements. We will provide a complete and formal description of the updated representation in \S\ref{section:grammar} and describe the API of \selfieslib in \S\ref{section:design}.

\section{SELFIES Specification} \label{section:grammar}

Being 100\% robust, every string of \SELFIES symbols corresponds to a \SMILES string that is both syntactically and semantically valid. Recall that we call a \SMILES string semantically valid if it syntactically valid and represents a molecular graph that obeys normal chemical valences. 

Within \SELFIES, these chemical valences are encoded as a constraint function $\valence \colon \mathcal{A} \to \mathbb{N}_0$, where $\mathcal{A}$ is a finite universe of the atom types (e.g., $\mathcal{A} = \{\textsf{C}, \textsf{N}, \textsf{O}, \textsf{F}, \ldots\}$) of interest and $\mathbb{N}_0 = \mathbb{N} \cup \{0\}$. The valences represented by $\valence$ dictate that an atom $A$ must assume $\valence(\atomtype(A))$ incident bonds in total. Note that if a \SMILES string obeys the valences $k$, each of its atoms $A$ makes \textit{at most} $\valence(\atomtype(A))$ explicit bonds within the string. There is a possibly-strict inequality in this case due to the way \SMILES automatically adds implicit hydrogens until chemical valences are satisfied. In practice, the mapping $\valence$ is rationally chosen to align with physical considerations and established cheminformatics packages such as RDKit \cite{landrum2006rdkit}. For example, a plausible setting might map 
\begin{equation}\label{eq:concrete constraints}
\valence(\textsf{C}) = 4, \quad \valence(\textsf{N}) = 3, \quad \valence(\textsf{O}) = 2, \quad \valence(\textsf{F}) = 1  
\end{equation}
which is the default behaviour of \selfieslib (see \S\ref{subsec:customization}).  

We formulate chemical valences in this manner to emphasize that although \SELFIES depends on $\valence$, it is not fixed to any particular setting of $\valence$. That is to say, \SELFIES can enforce rule sets induced by any arbitrary mapping $\valence \colon \mathcal{A} \to \mathbb{N}_0$, even if they are not chemically meaningful. To highlight an absurd example, the uniform constraints $\valence(\cdot) = 1000$ can be used in principle, which corresponds to effectively having no semantic constraints at all. In this sense, \SELFIES can be thought of as a general framework for an adjustable set of constraints $\valence$. In the ensuing discussion, we will describe \SELFIES under the assumption that some constraint function $\valence$ is fixed beforehand.

\subsection{Syntax}
 
Before explaining the \SELFIES specification, we make a brief aside and give an overview of the form of \SELFIES strings. Simply, a valid \SELFIES string is \textit{any} finite sequence of \SELFIES symbols joined together. For ease of visual partitioning, all \SELFIES symbols are enclosed by square brackets. Hence, a generic \SELFIES string is of the form     
\begin{equation}
    \code{[...][...]} \cdots \code{[...][...]}
\end{equation}
where the $\texttt{...}$ is a placeholder for a symbol-specific token. We can further categorize \SELFIES symbols into four main types, namely, atom, ring, branch, and miscellaneous, and characterize the syntax of each in the following. Throughout, let $\varepsilon$ be the empty string and given $n$ strings $(\sigma_i)_{i = 1}^n$, let $\sigma_1 \concat \sigma_2 \concat \cdots \concat \sigma_n$ denote their concatenation.

\vspace{\baselineskip}
 
\textbf{Atom Symbols.} The general \SELFIES atom symbol has the form 
\begin{equation}\label{eq:atom symbol form}
\begin{gathered} 
    \code{[} \concat \beta \concat \alpha \concat \code{]} \\
    \alpha = \alpha_{\text{iso}} \concat \alpha_{\text{elem}} \concat  \alpha_{\text{chiral}} \concat \alpha_{\text{H}} \concat \alpha_{\pm}
\end{gathered}
\end{equation}
where $\beta \in \{\varepsilon, \code{=}, \code{\#}, \code{/}, \code{\textbackslash}\}$ is a \SMILES-like bond symbol and 
\begin{align}
    \begin{split}
    \alpha_{\text{iso}} &\in \{\varepsilon, \code{1}, \code{2}, \code{3}, \ldots \} \\
    \alpha_{\text{elem}} &\in \{\text{element symbols}\}\\
    \alpha_{\text{chiral}} &\in \{\varepsilon, \code{@}, \code{@@}\}\\
    \alpha_{\text{H}} &\in \{\varepsilon, \code{H0}, \code{H1}, \ldots, \code{H9}\} \\
    \alpha_{\pm} &\in \{\varepsilon, \code{+1}, \code{-1}, \code{+2}, \code{-2}, \code{+3} \ldots\}
    \end{split}
\end{align}
collectively specify an atom type $\atomtype(\alpha)$ in a \SMILES-like fashion (the atom's isotope number, atomic number, chirality, number of attached hydrogens, and charge, respectively, and sometimes optionally). Notably, each \SELFIES atom symbol is semantically unique, i.e., different atom symbols are not interchangeable. This is not the case in \SMILES due to shorthand abbreviations in how attached hydrogens and charge can be represented. For example, the \SMILES atom symbol pairs $(\code{[Fe++]}, \code{[Fe+2]})$ and $(\code{[CH]}, \code{[CH1]})$ are interchangeable. To create a more standardized alphabet of symbols, we remove this redundancy in \SELFIES. 

\vspace{\baselineskip}

\textbf{Branch Symbols.} The general \SELFIES branch symbol has the form
\begin{equation} \label{eq:branch symbol form}
    \code{[} \concat \beta \concat \code{Branch} \concat \ell \concat \code{]}
\end{equation}
where $\beta \in \{\varepsilon, \code{=}, \code{\#}\}$ is a \SMILES-like bond symbol and $\ell \in \{\code{1}, \code{2}, \code{3}\}$. 

\vspace{\baselineskip}

\textbf{Ring Symbols.} \SELFIES ring symbols can be further subdivided into two sub-types. These are of the form 
\begin{equation}\label{eq:ring symbol form}
\begin{gathered}
    \code{[} \beta \concat \code{Ring} \concat \ell \concat \code{]} \\
    \code{[} \concat \beta_1 \concat \beta_2 \concat \code{Ring} \concat \ell \concat \code{]}
\end{gathered}
\end{equation}
where $\beta \in \{\varepsilon, \code{=}, \code{\#}\}$ and 
\begin{equation}
    \beta_1, \beta_2 \in \{\code{-}, \code{/}, \code{\textbackslash}\}, \text{ \textit{not} both } \beta_1 = \beta_2 = \code{-}  
\end{equation}
are \SMILES-like bond symbols and $\ell \in \{\code{1}, \code{2}, \code{3}\}$, similar to branch symbols. The second ring symbol type (Eq.~\ref{eq:ring symbol form}) is used to handle stereochemical specifications across double ring bonds (see \S\ref{subsec:ring derivation})

\vspace{\baselineskip}

\textbf{Miscellaneous Symbols.} \SELFIES has a few auxiliary symbols that are not core to the representation. These symbols still have common use cases and are specially recognized by the functions in \selfieslib that translate between \SELFIES strings and \SMILES strings (see \S\ref{subsec:core}):
\begin{itemize}
    \item The dot symbol \code{.}, which can be used to express multiple disconnected fragments in a single \SELFIES string, similar to its role in \SMILES. The dot symbol is interpreted by treating it as delimiter and splitting the \SELFIES string across the symbol. Then, each token is treated as an independent \SELFIES string.    
    
    \item The \code{[nop]} (for ``no-operation'') symbol, which is a special padding symbol ignored by \selfieslib.
\end{itemize}

\begin{table}[htbp]
    \caption{Example \SELFIES symbols, by symbol type.}
    \vspace{\baselineskip}
    \centering
    \begin{tabular}{lc}
        \toprule
         Type & Examples  \\
         \midrule
         Atom & \code{[\#13C]}, \code{[=O]}, \code{[C@@H1]}, \code{[N+1]}  \\
         Branch & \code{[Branch3]}, \code{[\#Branch1]}, \code{[=Branch2]} \\
         Ring & \code{[=Ring1]}, \code{[/\textbackslash{}Ring3]}, \code{[Ring2]} \\
         Misc. & \code{.}, \code{[nop]} \\
         \bottomrule
    \end{tabular}
    \label{tab:example symbols}
\end{table}

\autoref{tab:example symbols} provides examples of \SELFIES atom, branch, and ring symbols.

\subsection{The \SELFIES Grammar}

Now, we return to explaining the practical algorithm used to derive \SMILES strings from their correspondent \SELFIES strings. To do so, we first introduce the notion of a context-free grammar. A context-free grammar $G$ is a tuple $G = (V, \Sigma, R, S)$, where $V$ and $\Sigma$ are disjoint finite sets of nonterminal and terminal symbols, respectively, $R \subseteq V \times (V \cup \Sigma)^*$ is a finite relation, and $S \in V$ is a so-called start symbol. Under $G$, strings of terminal symbols can be derived by performing a finite sequence of replacements starting with the single-symbol string $\sigma_0 = S$. At each step $t$, if the current string $\sigma_t$ contains a nonterminal symbol $\mathbf{A} \in V$ (i.e., $\sigma_t = \rho_1 \concat \mathbf{A} \concat \rho_2$ for $\rho_1, \rho_2 \in (V \cup \Sigma)^*$) and there is an $(\mathbf{A}, \alpha) \in R$, then we replace $\mathbf{A}$ with $\alpha$ to get the next string $\sigma_{t + 1} = \rho_1 \concat \alpha \concat \rho_2$. For this reason, tuples $(\mathbf{A}, \alpha) \in R$ are called production rules, and are suggestively notated $\mathbf{A} \to \alpha$. The derivation terminates once only terminal symbols remain. The derivation of \SMILES strings under \SELFIES is similar to the preceding process. In fact, a context-free grammar underlies \SELFIES, which we call the \SELFIES grammar.

Specifically, the \SELFIES grammar takes
\begin{align}
\begin{split}
V &= \{\bS, \bX_1, \bX_2, \bX_3, \ldots, \bX_{\max \valence(\mathcal{A})}\} \\
\Sigma &= \{\text{\SMILES symbols, e.g., \code{C}, \code{=}, \code{(}, $\ldots$}\}\\
S &= \bS
\end{split}
\end{align}
where $\max \valence(\mathcal{A})$ is the maximum valence of all atom types. The production rules $R$ will be characterized later. Given a \SELFIES string, its corresponding \SMILES string is then derived through a trajectory of replacements starting from $\bS$, as previously described. However, there are two further modifications that provides \SELFIES its strong robustness. First, the replacements that are performed are not chosen arbitrarily, but are instead dictated by the \SELFIES string of interest. At each derivation step, the next symbol of the \SELFIES string is read off and fully specifies which production rule is applied. We systematically design this symbol-to-rule mapping such that the final derived \SMILES string will always be valid. Second, \SELFIES augments the grammar with self-referencing functions. These self-referencing functions manipulate the derivation process in more complicated ways than simple replacements, so they are not production rules. However, as before, the manner in which these self-referencing functions are applied is also dictated by the symbols in the \SELFIES string. Thus, a \SELFIES string can be viewed as a recipe of instructions (the symbols) that guides string derivation under the \SELFIES grammar. 

\subsection{Simple Chain Derivation} \label{subsec:main chain derivation}

Herein, we begin by considering the simplest type of \SELFIES strings, those that correspond to simple chains of atoms. In \SMILES, simple chains of atoms are represented by sequences of alternating atom and bond \SMILES symbols, the latter of which can sometimes be left implicit by convention. Examples of such \SMILES strings include \code{CCCC} (n-butane) and \code{O=C=O} (carbon dioxide). Analogously, in \SELFIES, simple chains are represented by sequences of \SELFIES atom symbols, which can be understood as playing a similar role as a grouping of a \SMILES atom symbol and its preceding \SMILES bond symbol. Simple chains are the easiest to derive in \SELFIES, because the process occurs only through mere replacements, as in regular context-free grammars. 

The derivation of a simple chain starts with the initial string $\sigma_0 = \mathbf{S}$. Recall that the \SELFIES symbols dictate how production rules are applied. For simple chains, this is achieved by having each pair of \SELFIES atom symbol and nonterminal symbol $\mathbf{A} \in V$  determine a production rule of the form $\mathbf{A} \to \alpha \concat \mathbf{A}'$, where $\alpha \in \Sigma^*$ is a terminal string and $\mathbf{A}' \in V \cup \{\varepsilon\}$. Then, a sequence of replacements is iteratively performed by treating the \SELFIES string as a queue $\mathcal{Q}$ of \SELFIES symbols. At each step, the head of $\mathcal{Q}$ is popped\footnote{To \textit{pop} or \textit{dequeue} the head of a queue $\mathcal{Q}$ means to fetch and then remove the oldest item in $\mathcal{Q}$.} and, with a nonterminal symbol in the current string $\sigma_t$, is used to select and apply a production rule to get the next string $\sigma_{t + 1}$. Note that $\sigma_0 = \mathbf{S}$ is itself a single nonterminal symbol, and each rule induced by a \SELFIES atom symbol replaces one nonterminal symbol by another. Hence, throughout the derivation, the current string $\sigma_t$ will always contain at most one nonterminal symbol and there is never any ambiguity as to how or which production rule is applied. Once the current string has only terminal symbols or $\mathcal{Q}$ is empty, the process ends (since \SELFIES strings are finite, termination necessarily occurs). The final derived \SMILES string is read off by dropping all nonterminal symbols.

We now fully enumerate the \SELFIES atom symbol to production rule mapping. Let $\atomsymb$ be a generic atom symbol, as described in Eq.~\ref{eq:atom symbol form}. Based on this symbol, we first define the terminal string 
\begin{equation}
    \tilde{\alpha} = 
    \begin{cases}
    \alpha, &\text{if } \alpha \in \mathcal{O} \\ 
    \code{[} \concat \alpha \concat \code{]}, &\text{otherwise}
    \end{cases}
\end{equation}
where $\mathcal{O} = \{\code{B}, \code{C}, \code{N}, \code{O}, \code{S}, \code{P}, \code{F}, \code{Cl}, \code{Br}, \code{I}\}$ are the symbols of elements in the \SMILES organic subset. The string $\tilde{\alpha}$ can be thought of as transforming $\alpha$ into the \SMILES syntax. Then $\atomsymb$ together with the nonterminal symbol $\mathbf{S} \in V$ specifies the production rule:
\begin{equation}\label{eq:prod rule from S}
    \mathbf{S} \to \tilde{\alpha} \concat \mathbf{X}_{\ell}
\end{equation}
where $\ell = \valence(\atomtype(\alpha))$ is the valence of the atom type specified by $\alpha$, and  we hereafter define $\mathbf{X}_0 = \varepsilon$ to be the empty string to handle the case where $\ell = 0$. The atom symbol $\atomsymb$ together with the symbol $\mathbf{X}_i \in V$, where $1 \leq i \leq \max \valence(\mathcal{A})$, specifies a production of the form: 
\begin{equation}\label{eq:prod rule from Xi}
    \mathbf{X}_i \to \begin{cases}
    \beta_{\downarrow}(d_0) \concat \tilde{\alpha} \concat \mathbf{X}_{\ell - d_0}, &\text{if } \ell > 0 \\
    \varepsilon, &\text{if } \ell = 0
    \end{cases} 
\end{equation}
where $d_0 = \min(\ell, i, d(\beta))$. Here, $d(\beta)$ is a function that returns the order of the bond type represented by $\beta$:
\begin{equation}\label{eq:bond order fn}
    d(\beta) = \begin{cases}
    1, & \text{if } \beta \in \{\varepsilon, \code{/}, \code{\textbackslash}\}\\
    2, & \text{if } \beta = \code{=} \\
    3, & \text{if } \beta = \code{\#}
    \end{cases}
\end{equation}
and $\beta_{\downarrow}(n)$ is a function that demotes $\beta$ into a \SMILES token representing a bond of lower order $n \leq d(\beta)$:
\begin{equation}\label{eq:downgrade bond fn}
\beta_{\downarrow}(n) = \begin{cases}
\beta, & \text{if } d(\beta) = n  \\
\varepsilon, & \text{if } d(\beta) \neq n = 1 \\
\code{=}, &  \text{if } d(\beta) \neq n = 2 \\
\end{cases}
\end{equation}
In Eq.~\ref{eq:prod rule from S} and Eq.~\ref{eq:prod rule from Xi}, the nonterminal symbols $\mathbf{X}_m$ are intuitively \textit{memorizing} the maximum number of bonds that the most recently derived atom can adopt; the nonterminal symbol $\mathbf{X}_m$ can be understood as encoding that the last atom can make at most $m$ bonds. When the next atom is derived, the bond connecting it to
the preceding atom has its order decreased minimally such that the bond constraints are always satisfied. 

\vspace{\baselineskip}

\textbf{Example.} To show these production rules in a concrete setting, we will translate the \SELFIES string 
\begin{equation}
\mathcal{Q} = \code{[=C][O][\#C][F][C]}    
\end{equation}
along with the constraints in Eq.~\ref{eq:concrete constraints}. The derivation of its corresponding \SMILES string would proceed step-wise as follows:
\begin{equation}
\begin{aligned}
   \qquad \qquad  \mathbf{S} &\implies \code{C} \concat \mathbf{X}_4  &&(\code{[=C]}) \qquad  \qquad  \\
    &\implies \code{CO}\concat \mathbf{X}_1  &&(\code{[O]}) \\
    &\implies \code{COC}\concat \mathbf{X}_3  &&(\code{[\#C]}) \\
    &\implies \code{COCF} \concat \varepsilon  && (\code{[F]}) \\
    &\implies \text{done.}
\end{aligned}
\end{equation}
where each line $\sigma_t \implies \sigma_{t + 1} \, (\atomsymb)$ is used to denote a step of the derivation process induced by the \SELFIES symbol $\atomsymb$. The final derived \SMILES string in this case is \code{COCF}.

\subsection{Branch Derivation} \label{subsec:branch derivation}
So far, we discussed chains of atoms, and their connectivity. However, most molecules are more complex than simple linear chains. Therefore, now, we talk about the derivations of branches (followed by rings in the subsequent section). In \SMILES, branches are specified by enclosing a \SMILES substring in parentheses, which can be recursively nested; for example, \code{CC(=O)O} (acetic acid) and \code{C(=O)(C(=O)O)O} (oxalic acid). In \SELFIES, branches are specified by \SELFIES branch symbols, and similar to atom symbols, every pair of \SELFIES branch symbol and nonterminal symbol determine some rule on how to modify the current string. We can encode branched trees of atoms in \SELFIES by sequences of atom and branch symbols.  

The derivation process extends that for simple chains (in \S\ref{subsec:main chain derivation}), where we pop \SELFIES symbols step-by-step off of a queue $\mathcal{Q}$. We only add an additional rule for when we dequeue a branch symbol from $\mathcal{Q}$. Let this symbol be $\branchsymb$, as in Eq.~\ref{eq:branch symbol form}, and let $\mathbf{A}$ be a nonterminal symbol in the current string $\sigma_t$. If $\mathbf{A} \in \{\mathbf{S}, \mathbf{X}_1\}$, then this specifies the application of the production rule $\mathbf{A} \to \mathbf{A}$.
Effectively, the branch symbol is ignored in this case. If $\mathbf{A} = \mathbf{X}_i$ for $i \geq 2$, then we perform a replacement:
\begin{equation}\label{eq:branch replacement}
    \mathbf{A} \to  \rho \concat  \mathbf{X}_{i - d_0}
\end{equation}
where $d_0 = \min(i - 1, d(\beta))$, and $\rho \in \Sigma^*$ is a \SMILES substring obtained through the following recursive process. 

\begin{table}[htbp]
	\caption{The symbols succeeding a branch or ring \SELFIES symbol are sometimes overloaded with a numeric index, which is determined by the following symbol-to-index mapping.}
	\vspace{\baselineskip}
	\centering
	\renewcommand*{\arraystretch}{1.3} 
	\begin{tabular}{cl|cl}
	\toprule
	Index & Symbol               & \; Index & Symbol               \\ 
	\midrule
	0     & \code{[C]}         & \; 8     & \code{[\#Branch2]} \\
	1     & \code{[Ring1]}     & \; 9     & \code{[O]}         \\
	2     & \code{[Ring2]}     & \; 10    & \code{[N]}         \\
	3     & \code{[Branch1]}   & \; 11    & \code{[=N]}        \\
	4     & \code{[=Branch1]}  & \; 12    & \code{[=C]}        \\
	5     & \code{[\#Branch1]} \; & \; 13    & \code{[\#C]}       \\
	6     & \code{[Branch2]}   & \; 14    & \code{[S]}         \\
	7     & \code{[=Branch2]}  & \; 15    & \code{[P]}         \\
	\midrule
	\multicolumn{4}{l}{All other symbols are assigned index 0.}  \\ \bottomrule
	\end{tabular}
	\label{tab:index alphabet}
\end{table}

First, $\ell$ symbols are popped from $\mathcal{Q}$ and converted into integer values by the mapping summarized in \autoref{tab:index alphabet}. Let $c_1 \cdots, c_\ell$ be the indices in first-to-last order of retrieval. In the event that $\mathcal{Q}$ contains fewer than $\ell$ symbols, the missing indices are set to have a default value of $0$. Next, these indices are identified with a natural number $N \in \mathbb{N}$ by treating them as hexadecimal digits: 
\begin{equation}\label{eq:hex index}
    N = 1 + \sum_{k = 1}^\ell 16^{\ell - k} c_k
\end{equation}
Then, $N$ symbols from $\mathcal{Q}$ (or all symbols in $\mathcal{Q}$,  if fewer exist) are consumed to form a new \SELFIES string, and with start symbol $S = \mathbf{X}_{d_0}$ (instead of $S = \mathbf{S}$ as before), this substring is recursively derived into a \SMILES string $\rho_0$. We take $\rho = \varepsilon$ if $\rho_0 = \varepsilon$, and $\rho = \code{(} \concat \rho_0 \concat \code{)}$ otherwise.\footnote{A minor technicality occurs if $\rho_0$ starts with a branch parentheses $\code{(}$, in which case $\rho$ is of the form $\code{((}  \alpha_1 \code{)} \cdots \code{(} \alpha_m \code{)} \alpha_{m + 1} \code{)}$ for strings $\alpha_k \in \Sigma^*$ that do not start with \code{(}. This would result in an invalid \SMILES string because branches cannot start with other branches in \SMILES. To amend this, we naturally interpret and replace $\rho$ with the string $\code{(}  \alpha_1 \code{)} \cdots \code{(} \alpha_m \code{)}\code{(}\alpha_{m + 1} \code{)}.$   
} 

\vspace{\baselineskip}

\textbf{Example.} To provide an overview of branch derivation, we translate a \SELFIES string representing acetic acid:
\begin{equation}
    \mathcal{Q} = \code{[O][C][=Branch1][C][=O][=C]}
\end{equation}
Processing the first two \SELFIES symbols $\code{[O][C]}$ results in the string $\code{OC}\concat \mathbf{X}_3$, after which the symbol $\code{[=Branch1]}$ is dequeued. Since $\ell = 2$, we consume the next symbol $\code{[C]}$ in $\mathcal{Q}$ and identify it with $N = 1$. Hence, we create the \SELFIES substring $\code{[=O]}$ from popping the next symbol in $\mathcal{Q}$ and, with start symbol $\mathbf{X}_2$, recursively derive it into the \SMILES substring $\rho = \code{(=O)}$. Then, performing the replacement in Eq.~\ref{eq:branch replacement} gives the string $\code{OC(=O)}\concat\mathbf{X}_1$, and processing the last symbol \code{[=C]} in $\mathcal{Q}$ finally produces a \SMILES string $\code{OC(=O)C}$ for acetic acid.

\subsection{Ring Derivation} \label{subsec:ring derivation}

The final feature that is necessary to capture the diverse variety of molecules is the ability to encode ring closures. In \SMILES, this is achieved by paired numeric tags that indicate two separate atoms are joined together; for example, \code{CC1CCC1} (methylcyclobutane). By adding bond characters before the numbers, \SMILES can also specify ring closures of higher bond orders, such as \code{C=1CCCC=1} (cyclopentene). 
In \SELFIES, ring closures are specified by ring symbols, which behave similarly to branch symbols. The derivation process extends that in \S\ref{subsec:branch derivation}. 

Per Eq.~\ref{eq:ring symbol form}, there are two forms of \SELFIES ring symbols. To simplify the ensuing discussion, however, we will begin by only considering the first form. When a ring symbol $\ringsymb$ is popped from the queue of \SELFIES symbols $\mathcal{Q}$, a nonterminal symbol $\mathbf{A}$ in the current derived string is used to specify a production rule. If $\mathbf{A} = \mathbf{S}$, then we apply the rule $\mathbf{A} \to \mathbf{A}$, and the ring symbol is effectively skipped. If $\mathbf{A} = \mathbf{X}_i$, then we replace: 
\begin{equation}
    \mathbf{A} \to \mathbf{X}_{i - \min(i, d(\beta))}
\end{equation}
In addition, we consume the next $\ell$ symbols of $\mathcal{Q}$ (or all symbols in $\mathcal{Q}$, if fewer exist) to specify a number $N \in \mathbb{N}$ by Eq.~\ref{eq:hex index}. Then, the ring symbol would indicate that a ring closure should be formed between the \textit{ring-initiating} atom and the $N$-th atom previously derived from it (or simply, the first atom if less than $N$ such atoms exist). Here, the  derivation order is the order in which atoms are realized through the production rules in Eqs.~\ref{eq:prod rule from S}~and~\ref{eq:prod rule from Xi}. By ring-initiating atom, we also mean the atom at which bonds would be made if the ring symbol were instead an atom symbol. Often, this coincides with the last-derived atom, as is the case in: 
\begin{equation}
    \code{NC(C)COC}^{\ast\dagger} \concat\mathbf{X}_4
\end{equation}
where the ring-initiating and last-derived atoms are marked with an asterisk and dagger, respectively. However, this is not the case when the last-derived atom lies within a fully-derived branch:
\begin{equation}
    \code{NC(C)COC}^\ast\code{(C)(C}^\dagger\code{)} \concat \mathbf{X}_1
\end{equation}
For brevity, we will refer to the ring-initiating atom as the \textit{right ring} atom and its counterpart the \textit{left ring} atom, as the latter precedes the former in a \SMILES string
under derivation order.  

Although a ring symbol specifies a closure between the left and right ring atoms, such a bond cannot be naively added since it may cause valences to be violated for the left ring atom immediately (e.g., consider the case where this atom has already attained its maximum valence) or in the future. 
Hence, \SELFIES postpones the creation of ring closures to a final post-processing step. Instead, the ring closure candidates are pushed to a temporary queue $\mathcal{R}$, and once all the \SELFIES symbols have been processed, the items in $\mathcal{R}$ are revisited in first-to-last order. Based on the state of the ring atoms, a candidate may be rejected (and no ring bond is made) or executed.

Specifically, given a potential ring closure indicated by symbol $\ringsymb$, let $m_1$ and $m_2$ be the number of additional bonds that the left and right ring atoms can make, respectively. If $m_1 = 0$ or $m_2 = 0$, we must reject the candidate since adding the ring closure would exceed one of the valences of the ring atom. The candidate is also rejected if its left and right ring atoms are not distinct, to avoid unphysical self-loops. Otherwise, the candidate is accepted, and, assuming there is no pre-existing bond between its two ring atoms, we form a new bond of order $d_0 = \min(d(\beta_1), m_1, m_2)$ between them. If a prior bond does exist (e.g., if a duplicate ring closure is specified earlier in $\mathcal{R}$), then we increment the order of this existing bond as necessary. That is, if the existing bond is of order $d_1$, then we promote it to a bond of potentially-higher order $\min(3, d_1 + d_0)$. 

\vspace{\baselineskip}

\textbf{Example.} We translate a \SELFIES string representing methylcyclobutane: 
\begin{equation}\label{eq:ring example}
    \mathcal{Q} = \code{[C][C][C][C][C][Ring1][Ring2]}
\end{equation}
The first five symbols produce the string $\code{CCCCC}\concat\mathbf{X}_4$, after which the ring symbol \code{[Ring1]} is dequeued. Since $\ell = 1$, the next and final symbol $\code{[Ring2]}$ specifies a single ring bond between the final $\code{C}$ and its $N = 3$rd preceding atom. This produces the \SMILES string $\code{CC1CCC1}$.

\vspace{\baselineskip}

The second ring symbol form $\stereoringsymb$ in Eq.~\ref{eq:atom symbol form} behaves nearly identically to $\code{[Ring}\concat\ell\code{]}$, and is used to support specification of stereochemistry across single ring bonds. The only difference occurs when a ring closure candidate produced by $\stereoringsymb$ is accepted, and a new ring bond is added between the two ring atoms. In this case, if $\beta_1 \in \{\code{/}, \code{\textbackslash}\}$, then we add the bond character $\beta_1$ before the numeric ring tag on the left ring atom, and similarly with $\beta_2$ and the right ring atom. For example, if the example Eq.~\ref{eq:ring example} used the symbol \code{[/-Ring1]} instead of \code{[Ring1]}, then the derived \SMILES string would be \code{CC/1CCC1}.  
\section{Library Design}
\label{section:design}

The \selfieslib library is designed to be fast, lightweight, and user-friendly. A small but nice feature of \selfieslib is that it also requires no extra dependencies. At its core, there are two functions that facilitate the interconversion between \SELFIES strings and \SMILES strings. For more advanced usage, we provide functions to customize the underlying semantic constraints that \selfieslib enforces and operates upon. Finally, we also provide a variety of utility functions for manipulating \SELFIES strings. The following describes each type of function in more detail and provides potential use case examples. All code snippets are written in Python, with \selfieslib being a Python library.

\subsection{Core Functions} 
\label{subsec:core}

\SELFIES strings can conveniently be created from and turned into \SMILES strings using the functions \code{encoder()} and \code{decoder()}, respectively. The latter derives a \SMILES string from a \SELFIES string, using the procedure described in \S\ref{section:grammar}. The former performs the reverse translation such that passing a \SMILES string through the composition \code{decoder(encoder())} is always guaranteed to recover a \SMILES string that represents the same molecule (but not necessarily the original \SMILES string itself). The following excerpt defines a toy function \code{roundtrip()} that illustrates this: 

\begin{lstlisting}[language=Python]
import selfies as sf

def roundtrip(smiles): 
    try:
        selfies = sf.encoder(smiles)      
        return sf.decoder(selfies)  
    except sf.EncoderError:
        return None  
        
benzene = roundtrip("c1ccccc1")  
# -> [C][=C][C][=C][C][=C][Ring1][=Branch1]
# -> C1=CC=CC=C1
\end{lstlisting}

\noindent 
Line 5 translates the \SMILES string for benzene into the \SELFIES string in Line 11. Notably, \SELFIES does not support aromatic atom symbols (e.g., \code{c}) in the same way as \SMILES, so \code{encoder()} performs an internal kekulization if it is passed an aromatic \SMILES string. 
Line 7 guards against errors raised by \code{encoder()} when being passed \SMILES strings that are syntactically invalid, semantically invalid (i.e., violate the constraints described in the next subsection), or unsupported. An unsupported \SMILES string uses features of \SMILES that are not implemented in \SELFIES, such as the wildcard \code{*} and quadruple bond \code{\$} symbols; the API reference of \selfieslib further details which \SMILES strings are currently supported. Line 10 applies the \code{roundtrip()} function to a \SMILES string \code{c1ccccc1} for benzene. Indeed, this round-trip translation recovers a \SMILES string \code{C1=CC=CC=C1} that is different than the original string, but still specifies the (kekulized) benzene molecule. 

Since every string of \SELFIES symbols can be derived into a valid \SMILES string, we can generate random but valid \SMILES strings by passing random \SELFIES strings through \code{decoder()}. To sample these \SELFIES strings, we use the \code{get\_semantic\_robust\_alphabet()} utility function, which returns a subset of semantically constrained \SELFIES symbols:
\begin{lstlisting}[language=python]
import random 

length = 10
alphabet = sf.get_semantic_robust_alphabet()
alphabet = list(alphabet)

symbols = random.choices(alphabet, k=length)
random_selfies = "".join(symbols)
random_smiles = sf.decoder(random_selfies)
\end{lstlisting}
Note that by changing the pool of \SELFIES symbols from which we sample from, we can change the distribution of produced molecules.

\subsection{Explaining Translation} 

To explain translations between \SELFIES and \SMILES,  both \code{encoder()} and \code{decoder()} support an \code{attribute} flag that enables attributions of the output string symbol(s) to symbol(s) in the input string:

\begin{lstlisting}[language=Python]
cyclobutane = "[C][C][C][C][Ring1][Ring2]"
smiles, attributions = sf.decoder(
    cyclobutane, 
    attribute=True,
)

# smiles = C1CCC1
# attributions is a length-4 list with: 
attributions[0] = AttributionMap( 
    index=0, 
    token="C", 
    attribution=[
        Attribution(index=0, token="[C]")
    ],
)
attributions[1] = AttributionMap(
    index=2, 
    token="C", 
    attribution=[
        Attribution(index=1, token="[C]")
    ],
)
# attributions[2] = AttributionMap(...) 
# attributions[3] = AttributionMap(...)
\end{lstlisting}

\noindent
The attributions are a list of \code{AttributionMap} objects, one for each output symbol. Each \code{AttributionMap} contains the output symbol, its index, and a list of \code{Attribution} objects, each of which holds an input symbol (and its index) that is responsible for the output symbol. Note that a single output symbol may be attributed to multiple input symbols because it may be determined by both atom symbols and branch or ring symbols. Tracing the relationship between symbols can enable alignment between \SMILES and \SELFIES so that per-atom properties can be connected on both sides of the translation. 

\subsection{Customization Functions} 
\label{subsec:customization} 

\selfieslib dynamically constructs its derivation rules from a set of prespecified constraints, which dictate the maximum number of bonds that each atom type in a molecule may form. The derivation rules then ensure that each \SELFIES string corresponds to a molecular graph satisfying the set constraints. By choosing a set of constraints in accordance with chemical valences, 100\% robustness can be achieved. Specifically, \selfieslib uses the constraints in \autoref{tab:default constraints} by default.

\begin{table}[ht]
	\caption{The default constraints used by \selfieslib. All atom types other than those explicitly listed below are constrained to 8 maximum bonds, which acts as a \textit{catch-all} constraint.}
	\vspace{\baselineskip}
	\centering
	\begin{tabular}{cccc}
		\toprule
		& \multicolumn{3}{c}{Maximum Bonds} \\
		\cmidrule(r){2-4}
		Element     & Charge $0$ & Charge $+1$ & Charge $-1$ \\
		\midrule
		H, F, Cl, Br, I & 1 & / & / \\
		B & 3 & 2 & 4 \\
		C & 4 & 5 & 3 \\
		N & 3 & 4 & 2 \\
		O & 2 & 3 & 1 \\
		P & 5 & 6 & 4 \\
		S & 6 & 7 & 5 \\
		\bottomrule
	\end{tabular}
	\label{tab:default constraints}
\end{table}

However, a limitation of the default constraints is that \SELFIES cannot represent existing molecules that violate them, such as perchloric acid (which features a hypervalent Cl making 7 bonds). Moreover, the catch-all constraint may be too relaxed to ensure the validity of \SELFIES strings containing atom types outside those in \autoref{tab:default constraints} (e.g., Si, Se). Hence, users may wish to instead use custom constraints that are tailored to the \SELFIES strings being worked with. To this end, \selfieslib provides the key function \code{set\_semantic\_constraints()}. The following provides a minimal example:  

\begin{lstlisting}[language=Python]
import selfies as sf

constraints = {
    "C": 4, "C+1": 5, "C-1": 3,
    "?": 4  # catch-all
}

sf.set_semantic_constraints(constraints)

\end{lstlisting}

\noindent 
Here, the \code{constraints} dictionary encodes a set of custom constraints; specifically, explicit constraints on the neutral and $\pm 1$ charged variants of C (as in \autoref{tab:default constraints}) and a catch-all constraint (of 4 maximum bonds). Line 8 then sets \code{constraints} as the underlying semantic constraints that \selfieslib will operate under, which changes the subsequent behaviour of \code{encoder()} and \code{decoder()} appropriately. 
Note that the pre-existing constraints are fully replaced in Line 8; any constraint that is not explicitly specified in \code{constraints} would be thus removed.

For convenience, \selfieslib provides a couple of preset constraints to serve as templates that can be easily modified. These can be obtained as follows:

\begin{lstlisting}[language=Python]
c1 = sf.get_preset_constraints("default")
c2 = sf.get_preset_constraints("octet_rule")
c3 = sf.get_preset_constraints("hypervalent")
\end{lstlisting}

\noindent
The currently-set constraints can also be viewed by:

\begin{lstlisting}[language=Python]
curr_constraints = sf.get_semantic_constraints()
\end{lstlisting}

\begin{table*}[t]
	\caption{An overview of \selfieslib utility functions.}
	\vspace{\baselineskip}
	\centering
	\renewcommand*{\arraystretch}{1.3} 
	\begin{tabular}{cl}
	\toprule
	Function               & Description \\ 
	\midrule
	\code{len\_selfies()} & Computes the symbol length of a \SELFIES string. \\
	\code{split\_selfies()} & Tokenizes a \SELFIES string into its constituent symbols. \\
    \code{get\_alphabet\_from\_selfies()} & Extracts a minimal vocabulary from a dataset of \SELFIES strings. \\
    \code{selfies\_to\_encoding()}    & Converts a \SELFIES string into a label and/or one-hot encoding. \\
    \code{encoding\_to\_selfies()}    & Recovers a \SELFIES string from its label and/or one-hot encoding. \\
    \code{get\_semantic\_robust\_alphabet()} & Provides an alphabet of semantically-constrained \SELFIES symbols. \\
	\bottomrule
	\end{tabular}
	\label{tab:utility summary}
\end{table*}

\subsection{Utility Functions} 

\selfieslib provides a number of utility and convenience functions. Two basic utility functions are \code{len\_selfies()}, which computes the number of symbols in a \SELFIES string, and \code{split\_selfies()}, which tokenizes a \SELFIES string into an iterable of its constituent symbols: 

\begin{lstlisting}[language=Python]
import selfies as sf

selfies = "[F][=C][=C][#N]"
length = sf.len_selfies(selfies)  # 4
symbols = list(sf.split_selfies(selfies))  
# ["[F]", "[=C]", "[=C]", "[#N]"]
\end{lstlisting}

\noindent
Furthermore, \selfieslib includes functions to extract a vocabulary of symbols from a dataset of \SELFIES strings, and to convert \SELFIES strings into label or one-hot encodings. Consider the following example: 

\begin{lstlisting}[language=Python]
dataset = [
    "[C][O][C]",
    "[F][C]",
    "[C][C][O][C]",
]

alphabet = sf.get_alphabet_from_selfies(dataset)
alphabet.add("[nop]")
alphabet = list(sorted(alphabet))
# ["[C]", "[F]", "[O]", "[nop]"]

pad_to = max(sf.len_selfies(s) for s in dataset)
stoi = {s: i for i, s in enumerate(alphabet)}

dimethyl_ether = dataset[0]  # [C][O][C]

label, one_hot = sf.selfies_to_encoding(
    selfies=dimethyl_ether,
    vocab_stoi=stoi,
    pad_to_len=pad_to,  # 4
    enc_type="both",
)

# label = [0, 2, 0, 3]
# one_hot = [[1, 0, 0, 0], [0, 0, 1, 0], 
#            [1, 0, 0, 0], [0, 0, 0, 1]]
\end{lstlisting}

\noindent
Here, we are given a list \code{dataset} of \SELFIES strings. Line 7 uses a utility function of \selfieslib to extract the set \code{alphabet} of \SELFIES symbols that appear in the dataset, which is used in Line 13 to create a symbol to index mapping \code{stoi}. Next, lines 17-22 use another utility function \code{selfies\_to\_encoding()} to create a label and one-hot encoding of the first \SELFIES string in \code{dataset}. Under the hood, this function first pads the input string to length \code{pad\_to\_len} by appending to it sufficiently many copies of the symbol \code{[nop]} (for ``no-operation''), which is a special padding symbol in \selfieslib that is automatically ignored by \code{decoder()}. Then, the padded \SELFIES string is tokenized, and \code{stoi} is used to convert each of its symbols into integer labels and one-hot vectors. Since the padded \SELFIES string may now contain \code{[nop]}, this symbol must be added to \code{stoi}, which is done through Line 8. Lastly, the reverse encoding can be performed using the \code{encoding\_to\_selfies()} utility: 

\begin{lstlisting}[language=Python]
itos = {i: s for s, i in stoi.items()}

# recover [C][O][C][nop] from label encoding
recovered = sf.encoding_to_selfies(
    encoding=label, 
    vocab_itos=itos,
    enc_type="label",
}

sf.decoder(recovered)  # COC
\end{lstlisting}

\noindent
\autoref{tab:utility summary} summarizes the various utility functions introduced within this section.

\section{Results and Discussion}
\label{sec:results}

\selfieslib is quick and efficient in its translation, despite being implemented in pure Python. To demonstrate this, we provide some simple benchmarks of its core functions \code{encoder()} and \code{decoder()}. The following experiments were run on Google Colaboratory, which uses two 2.20GHz Intel(R) Xeon(R) CPUs. 

\vspace{\baselineskip}

\begin{figure*}[htbp]
\centering
    \centering
    {
    \setlength{\fboxrule}{0pt}
    \framebox[0.11\textwidth]{}
    \framebox[0.36\textwidth]{(a) Basic Alphabet}
    \framebox[0.08\textwidth]{}
    \framebox[0.36\textwidth]{(b) Filtered Alphabet}
    \framebox[0.04\textwidth]{}
    }\\
    \includegraphics[width=0.85\textwidth]{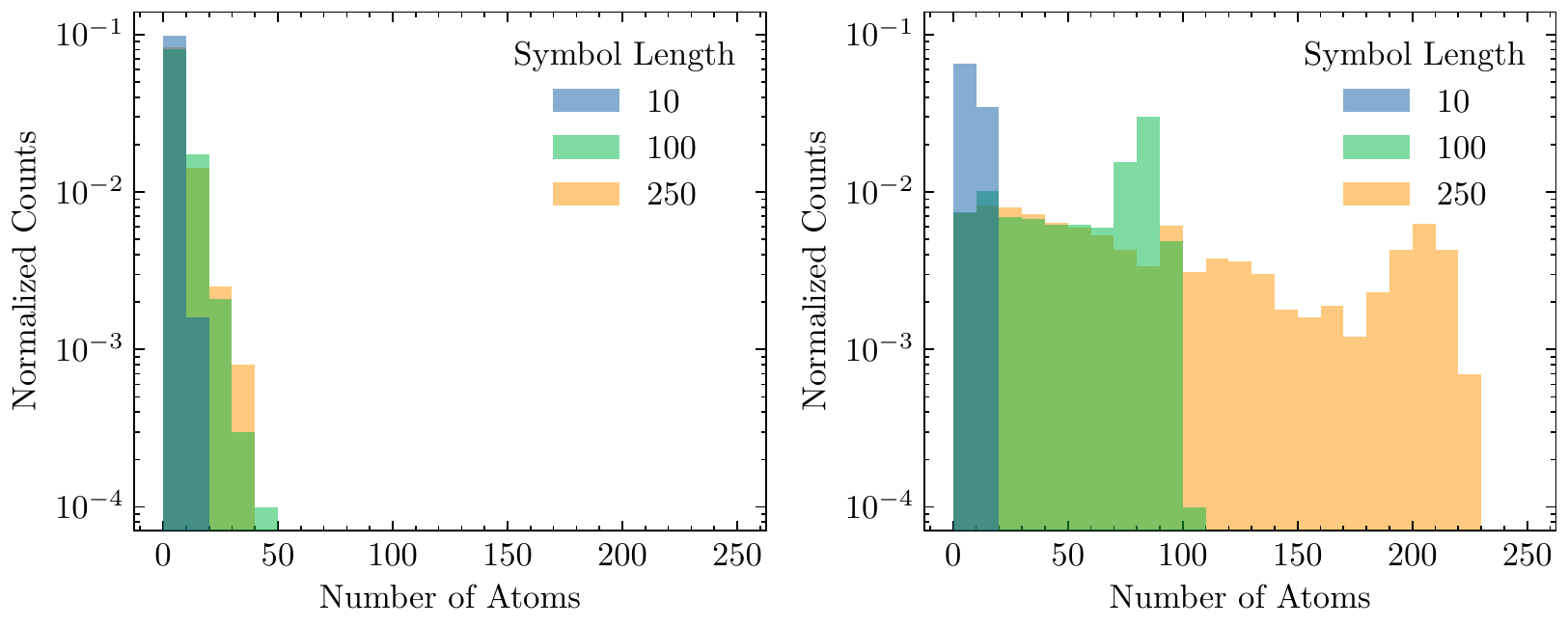} \\
    (c) Translation Speed \\
    \vspace{\baselineskip}
    \begin{tabular}{c|cc}
        \toprule
        Length \; & \; Basic & Filtered  \\
        \midrule
        10  & \; 0.082~s & 0.133~s \\
        100 & \; 0.199~s & 0.929~s \\
        250 & \; 0.341~s & 1.633~s \\
        \bottomrule
    \end{tabular}\\
    \caption{For a fixed alphabet $\mathcal{A}$, 1000 \SELFIES strings were generated by uniformly sampling $L$ symbols from an alphabet. Then, we plot the size distribution of the resulting molecules for varying symbol lengths $L$. \textbf{(a)} We take $\mathcal{A}$ to be the 69 symbols returned by \code{get\_semantic\_robust\_alphabet()} under the default semantic constraints. \textbf{(b)} We filter the alphabet in (a) to 19 symbols by removing all atom symbols $\atomsymb$ where $\beta \in \{\code{=}, \code{\#}\}$ or $\valence(\atomtype(\alpha)) = 1$, and removing all branch and ring symbols except for \code{[Branch1]} and \code{[Ring1]}. This decreases the chance that the \SELFIES derivation process is terminated early, causing the derived molecules to be larger. \textbf{(c)} The time taken to translate each batch of random \SELFIES strings to \SMILES using \code{decoder()}, measured by averaging over 20 replicate trials.
    }
    \label{fig:random selfies}
\end{figure*}

\begin{figure}[htbp]
    \centering
    \includegraphics[width=0.85\linewidth]{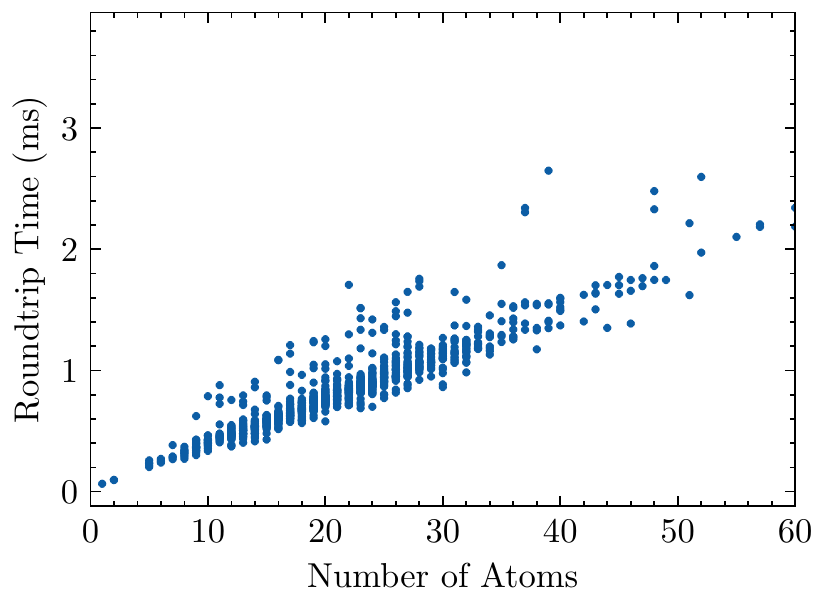}
    \caption{The roundtrip translation time of 1000 randomly-sampled \SMILES strings from the DTP open compound collection as a function of size, measured in number of atoms.}
    \label{fig:roundtrip translation}
\end{figure}

\textbf{Roundtrip Translation.} Here, we consider the roundtrip translation task, where a \SMILES string is translated to \SELFIES and then back to \SMILES (see \S\ref{subsec:core}). Specifically, we translate the Developmental Therapeutics Program (DTP) open compound collection \cite{voigt2001comparison, ihlenfeldt2002enhanced}, which contains a little over 300k \SMILES strings and is a set of molecules which have been tested experimentally for potential treatment against cancer and the acquired immunodeficiency syndrome (AIDS) \cite{milne1994national}. Translating the full dataset into \SELFIES strings with \code{encoder()} takes 136~s, and recovering the \SMILES dataset using \code{decoder()} takes 116~s, for a total roundtrip translation time of 252~s. 
\autoref{fig:roundtrip translation} plots how this roundtrip time scales with molecular size. Notably, we obtain all of these times by averaging over 3 replicate trials. 

\vspace{\baselineskip}

\textbf{Random \SELFIES.} First, we sample 1000 fixed-length \SELFIES strings and translate them to \SMILES, per \S\ref{subsec:core}. We try this experiment with different symbol lengths and alphabets from which the \SELFIES strings are built. \autoref{fig:random selfies} shows the resulting distribution of \SMILES strings and the time it takes to decode each full batch of random \SELFIES strings. Performing this experiment reaffirms the robustness of \SELFIES and demonstrates the ease in which we can create random valid molecules without applying any filters, pre- or post-selection. In \autoref{fig:random selfies}(a), we show how \SELFIES strings sampled from a basic alphabet translate to random molecules; an important observation is that the generated molecules are rather small, independent of the \SELFIES length chosen. That is mainly caused by the inclusion of multi-bonds and low-valence atoms in the considered alphabet, which exhaust the available valences of the constituent atoms and then lead to an earlier termination of the derivation. 
A simple workaround is to instead use an alphabet without multi-bonds and low-valence atom types, as illustrated in \autoref{fig:random selfies}(b). Here, the molecular size distribution is shifted significantly towards larger molecules, especially when longer \SELFIES string are sampled. Hence, this showcases how to create very large, valid random molecules.

\section{Conclusions and Outlook} \label{sec:conclusions}

Since its first release in 2019, the \selfieslib library has undergone significant changes and experienced a drastic transformation in terms of both capabilities and code design. All of these modifications were executed with two major premises, namely, (1) extending the functionality and capability to support all features of the \SMILES representation and (2) retaining or even improving upon its simplicity and user-friendliness. To achieve that, we implemented all necessary functionality in the library itself so that it does not require any other packages. Additionally, we added several utility functions to the library to support common use cases. Apart from these two prime goals, we also made significant efforts to make the implementation faster as \SELFIES has been employed in many performance-critical applications and workflows. 


Overall, the \SELFIES community has grown rapidly and we are actively engaging in constructive discussions about the current implementation and future improvements. While \selfieslib \latestversion supports almost all important features of \SMILES, there are still many new features on our agenda. We outlined many of them in \cite{krenn2022selfies}, for example, extensions to polymers, crystals, molecules with non-covalent bonds, or reactions. Our vision is that \SELFIES will become a standard computer representation for molecular matter. We encourage the community to implement it into their workflows, report errors in the current implementation, and propose changes and new features that will help them to succeed in their goals.

\bibliography{refs}

\end{document}